\documentclass[twocolumn]{sig-alternate-10pt}
\usepackage[sort,nocompress,space]{cite}

\usepackage{listings}
\usepackage{lineno}

\usepackage{pifont}
\usepackage{epsfig,epsf,url,amssymb}
\usepackage{tabularx}
\usepackage{float}
\usepackage[linesnumbered,boxed,ruled,vlined]{algorithm2e}
\usepackage{amsmath}
\usepackage{mathtools}

\usepackage{amsthm}
\usepackage{rotating}
\usepackage{times}
\long\def\comment#1{}
\usepackage{multirow}
\usepackage{lscape}
\usepackage{stmaryrd}
\usepackage{wrapfig}
\usepackage{hhline}
\usepackage{textcomp,booktabs}
\usepackage[usenames,dvipsnames]{color}
\usepackage{colortbl}
\usepackage[font=bf, labelfont=bf,skip=0pt]{caption}
\newcommand{\sys}{{\textsf{p4mr}}\xspace}

\long\def\comment#1{}

\newtheorem*{theorem-non}{Theorem}

\newenvironment{icompact}{
  \begin{list}{$\bullet$}{
    \parsep 0.5pt plus 0.5pt
    \partopsep 0.5pt plus 0.5pt
    \topsep 0.5pt plus 1pt minus 0.5pt
    \itemsep 0.5pt plus 0.5pt
    \parskip 0pt plus 1pt
    \leftmargin 0.15in}
       }
  {\normalsize\end{list}}
\newcommand{\paraspace}{\vspace{0.05in}} %0.05
\newcommand{\parab}[1]{\paraspace\noindent{\bf #1} }

\newcount\hour \newcount\minute
\hour=\time  \divide \hour by 60
\minute=\time
\loop \ifnum \minute > 59 \advance \minute by -60 \repeat
\def\drafttime{\ifnum \hour<13 \number\hour:%
                      \ifnum \minute<10 0\fi
                      \number\minute
                      \ifnum \hour<12 \ AM\else \ PM\fi
         \else \advance \hour by -12 \number\hour:%
                      \ifnum \minute<10 0\fi
                      \number\minute \ PM\fi}

\usepackage{epstopdf}

\usepackage{multirow}
\usepackage{rotating}

\usepackage{hyperref}
\hypersetup{
  colorlinks=true,      % false: boxed links; true: colored links
  linkcolor=blue,       % color of internal links
  citecolor=magenta,    % color of links to bibliography
  filecolor=cyan,       % color of file links
  urlcolor=red          % color of external links
}

\begin{document}
%
%\conferenceinfo{SIGCOMM,} {XXXXXX, XXXXX, XXXXX, XXXXX.} %
%\CopyrightYear{2009} \crdata{1-59593-308-5/06/0009}

\title{Programmable Switch as a Parallel Computing Device}
\numberofauthors{4} 
\author{
% 1st. author
\alignauthor
Li Chen\\
       \email{lchenad@ust.hk}
% 2nd. author
\alignauthor
Ge Chen\\
       \email{gchenaj@connect.ust.hk}
% 3rd. author
\alignauthor Justinas Lingys\\
       \email{jlingys@connect.ust.hk}
\and  % use '\and' if you need 'another row' of author names
% 4th. author
\alignauthor Kai Chen\\
\affaddr{SING Group}\\
\affaddr{CSE,HKUST}\\
       \email{kaichen@cse.ust.hk}
}
% There's nothing stopping you putting the seventh, eighth, etc.
% author on the opening page (as the 'third row') but we ask,
% for aesthetic reasons that you place these 'additional authors'
% in the \additional authors block, viz.
\date{\today}
% Just remember to make sure that the TOTAL number of authors
% is the number that will appear on the first page PLUS the
% number that will appear in the \additionalauthors section.

\maketitle

\setlength{\textfloatsep}{0pt}% Remove \textfloatsep
\medmuskip=0mu\relax % Remove spaces around $\times$
\thinmuskip=0mu\relax % Remove spaces around $\times$
\thickmuskip=0mu\relax % Remove spaces around $\times$
%\vspace{-0.05in}
\abstract{
Modern switches have packet processing capacity of up to multi-tera bits per second, and they are also becoming more and more programmable. We seek to understand whether the programmability can translate packet processing capacity to computational power for parallel computing applications. In this paper, we first develop a simple mathematical model to understand the costs and overheads of data plane computation. Then we validate the the performance benefits of offloading computation to network. Using experiments on real data center network, we find that offloading computation to the data plane results in up to $20\times$ speed-up for a simple Map-Reduce application. Motivated by this, we propose a parallel programming framework, \sys, to help users efficiently program multiple switches. We successfully build and test a prototype of \sys on a simulated testbed.

% \sys is a framework to allow applications to offload computation to the data plane. It enables Map-Reduce style data flow computation in the datacenter fabric. \sys enables information aggregation in the data plane, which reduces the traffic and lessons the load in servers.
}

\section{Introduction}
The undeniable trend in datacenter networks (DCN) is that the switches in the data plane are becoming more and more programmable~\cite{p416lang,p4,pof,smartnic}. With recent advances in switch architecture\cite{p4}, novel hardware component design\cite{pifo}, and programming language support\cite{domino,p4}, user-defined functions have been shown to perform on programmable switches at line rate of 10/40Gbps. 

With programmable switch, many applications have been proposed (switch-based key-value store~\cite{switchkv}, transport layer load balancing~\cite{hula}, in-band telemetry~\cite{int}, consensus algorithm~\cite{paxosp4}, etc.). Among all these applications, one of the guiding principle is to minimize the per-packet computation done on the switches. This \emph{minimal computation principle} (MCP) serves to reduce per-packet latency and overhead on the switches, to lower memory footprint, and to increase throughput.

The merits of MCP is easily recognized, and we concede that this principle should be followed in general for data plane applications. However, we cannot ignore the fact that the packet processing power is up to 6.5Tbps~\cite{tofino} for modern programmable switches. Tempted by such raw computational potential, in this paper, we aim to explore the opposite of MCP: we try to understand whether programmable switches can be used as a parallel computing device. Our insight is drawn from the success of using GPU to offload massively parallel tasks from CPU. We envision that (some) parallel computing tasks can be offloaded to the network from the servers. In other words, while the data is being transmitted in the network, we program the switches to perform some computation on the data, so that when the data reaches the destination, they are already processed.

%computation is plenty in existing data plane, e.g. counting, routing, forwards, and state machines. 
%They are done at line rate of the networking device. 
%These fixed data plane functions are traditionally distinct from user applications. 
% The emergence of programmable data plane blurs this distinction, and unlocks the possibility of doing user computation tasks at line rate in the data plane.

To this end, the following questions must be answered. 
\begin{icompact}
\item What are the primitive computations that can be supported by programmable switches?
\item What is the performance penalty of doing computing tasks on programmable switches?
\item What are benefits of doing computing tasks on programmable switches?
\item How to program a network of switches to perform computing tasks?
\end{icompact}

In this paper, we describe our initial efforts to allow programmable switches to support data plane computation, and provide some preliminary answers to the above questions.
In particular, we use the Word-Count application as a running example. First we describe the primitive functions (written in P4) that can be implemented with one programmable switches, and show that a network of programmable switches with different primitive functions can work together to complete the Word-Count application ($\S$\ref{sec:wordcount}). Then, we develop a mathematical model to understand the performance penalty of offloading computation to the network ($\S$\ref{sec:model}). We also use experiments to demonstrate the performance benefits of doing data plane computations ($\S$\ref{sec:expr}). Finally, we propose a parallel programming framework, \sys, that facilitates programming a network of programmable switches ($\S$\ref{sec:p4mr}).

\begin{figure*}[t]
\centering
  \includegraphics[width=\linewidth]{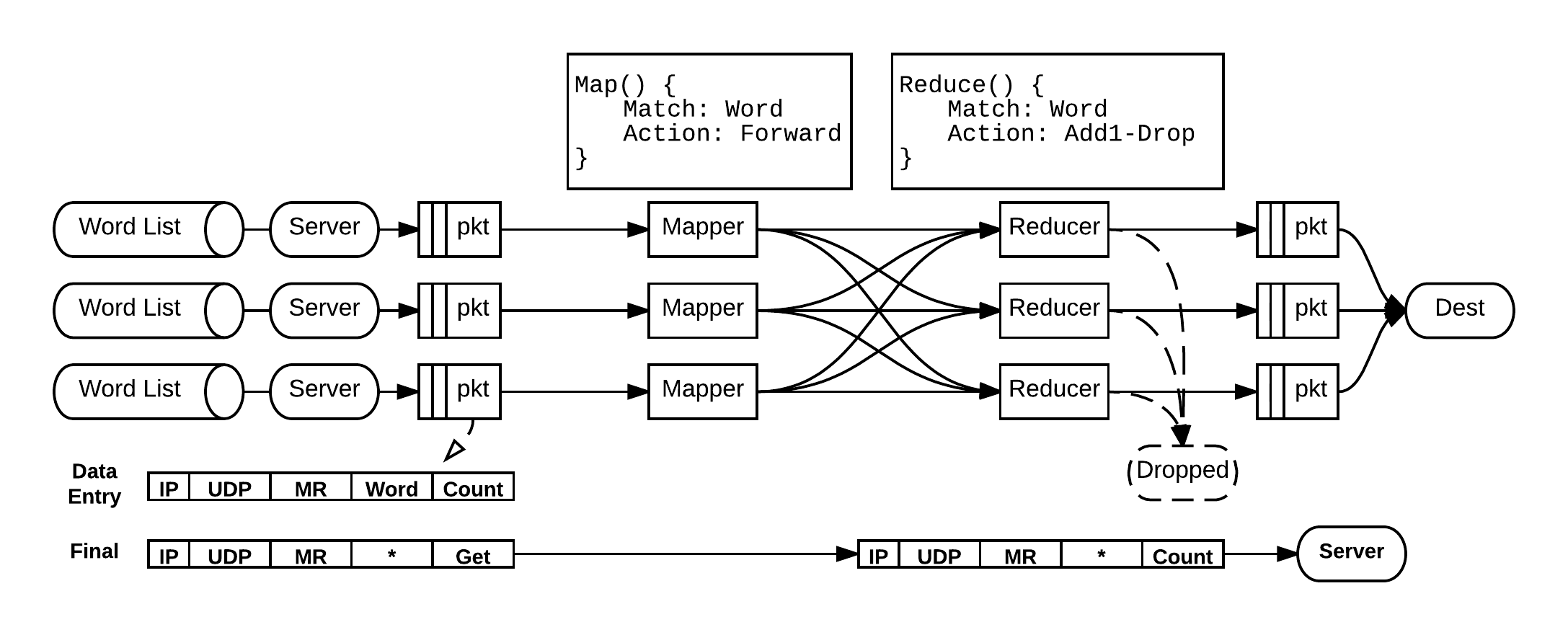}
  \caption{Example: Word-Count}\label{fig:wc}
\end{figure*}

\section{Offloading Word-Count to Data Plane}\label{sec:wordcount}
In this section, we use the well-known Word-Count applications to introduce computation in data plane. 

In a conventional Map-Reduce framework (Hadoop~\cite{yahoohadoop}, Spark~\cite{spark}, etc.), Word-Count runs as follows on multiple servers. Text files (or word lists) are located on different servers, and Map tasks are assigned to these servers. Map tasks maps each word ("alice", "bob", etc.) in the text file to tuples (<"alice", 1>, <"bob",1>, etc.). These tuples are send from the mappers (server running Map task) to the reducers (servers running Reduce tasks) based on their hash values. For the Reduce task, it accumulates the total appearance of words (<"alice", 5>, <"bob",1>, etc.). Each reducer is responsible for a subset of all the possible words. The final result of Word-Count is then the combination of all the results from all the reducers.

Due to the simplicity of Word-Count, we use it as an example of data plane computation. Suppose the data set of word lists are distributed among a set of servers.
Performing Word-Count using a network of programmable switches is very similar to that in a Map-Reduce framework. 
Servers \emph{serialize} their own word lists into stream of packets, and send them to mappers. The mapper is running on programmable switch. It processed each word, and forwards the word to a corresponding reducer based on the hash value of the word. The reducers are also running on a programmable switch, and it accumulates the count for each word. After a word is counted, it is discarded at the reducer.

However, we emphasize that there are some key differences from conventional Map-Reduce frameworks: Firstly, In the data plane, the smallest unit of computation is packet, and each packet should have a fixed format for the switch to process. Secondly, routing/forwarding based on matched field(s) is part of computation. Finally, to extract the final result, each server need to send a final packet after it has sent all the data packets, so as to trigger all the switches hosting the reducer to send to the destination hosts.

This example also reveals many unanswered design questions: 1) how should the servers serialize their data sets, or should serialization be also offloaded to the network? 2) should we do partial offload or full offload of Map and Reduce tasks to the network? 3) how to program multiple switches to perform the same job efficiently? In what follows, we attempt to provide some initial answers. 

%\subsection{Data plane compuation: fixed functions}
%There are many compuation functions that are already in fixed function switches.
%
%\subsection{Programmable data plane and P4}
%With P4, data plane becomes more and more programmable.

\vspace{0.3in}
\section{Cost of Data Plane Computation}\label{sec:model}
In this section, we look at the overhead of doing computation in the data plane. In particular, we examine the overhead of serialization --- the cost of transforming the data set into packets (each containing a single data item) that can be processed by the switch.

We make the following assumptions on the data set: 1) the data set contains multiple data items; 2) each data item is less than the MTU (maximum transmission unit~\cite{mtu}) size of the network; 3) each data entry have the same size.

The overhead of serialization mainly come from sending data packets. Given a data set, the servers can either 1) send each data item as an individual packet, or 2) send a MTU-sized packet packed with many data items\footnote{Only an integral number of data items can be packed into one packet.}. The difference between the choices is whether the server or the network should handle the overhead of serialization.

\begin{figure}[t]
\centering
  \includegraphics[width=\linewidth]{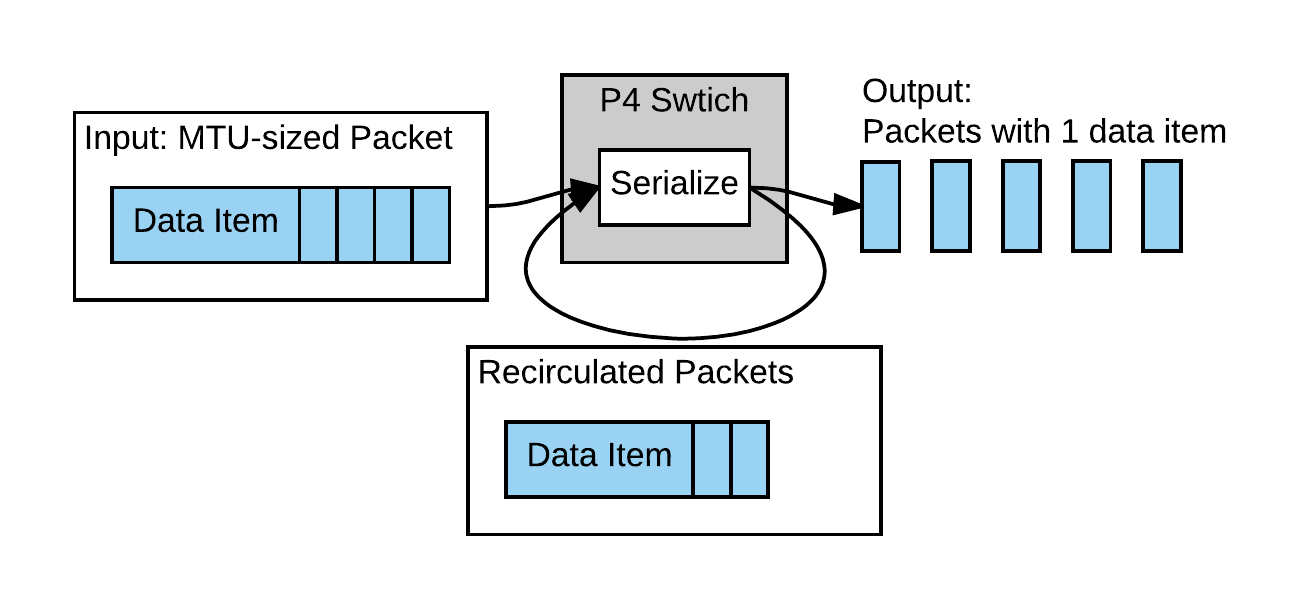}
  \caption{Serialization on the switch}\label{fig:serial}
\end{figure}

When serialization happens in the server, the main overhead is the CPU processing, and in the next section we use experiment to verify and compare the actual overhead. When the serialization happens in the switch, as shown in Figure~\ref{fig:serial}, each MTU-size packet is in itself a small data set containing multiple data items. For a packet containing $k$ items, to separate the items into $k$ different packets, the original packet must be recirculated~\cite{p416lang} $k$ times. In effect, this recirculation and separation process is analogous to the map task in Map-Reduce frameworks~\cite{hadoop}. Recirculation has throughput penalty on the switch undertaking this "map" task, and we model this penalty using a simple mathematical model.

Assume the server is sending MTU-size packets at rate $r$ to a switch port, and the switch is performing serialization. The maximum capacity of the switch is $C$. Since the switch keeps recirculating its input packets and at the same time taking in new packets at the rate $r$, we have the following equation at equilibrium:
\begin{equation}
  \lim_{N\rightarrow \infty} r(1+\frac{1}{N})^N = C
\end{equation}

This is because, if we divide the time in to $N$ slices before reaching equilibrium, in each slice, the input rate increase by $(1+\frac{1}{N})$. Therefore, we have $r=C/e$, and the penalty (loss) on throughput of a switch port with capacity C is $C(1-1/e)$.

Therefore, the decision to put serialization on server or on switch depend on which penalty (CPU or switch throughput) impact the performance the most. To answer this question, we perform a series of experiments on a real data center testbed.
\section{Benefits of Data Plane Computation}\label{sec:expr}
With penalties either on server CPU or switch throughput, one may wonder whether it is still beneficial to perform computational task in the data plane using switch. In this section, we use experiments to evaluate the usefulness of offloading computation to data plane.

We use the Word-Count example, which has two tasks (Map and Reduce, as shown in Figure~\ref{fig:wc}). The Map task serialize data sets on each server into data items, as we have described in $\S$\ref{sec:model}. The Reduce task aggregates the counts of different words. We design three scenarios:
\begin{icompact}
  \item \textbf{Scenario 1:} Map and Reduce are both done on servers.
  \item \textbf{Scenario 2:} Map is done on servers, and Reduce is done in the network. In this case, servers use CPU processing to separate data sets into data items, and send packets each containing one data item.
  \item \textbf{Scenario 3:} Map and Reduce are both done in the network. In this case, the servers send MTU-sized packets to the network, and each packets contains multiple data items. The switch recirculates the packets to generate new packets containing only one data item. Doing so incurs network throughput penalty of $1-1/e$, as shown in $\S$\ref{sec:model}.
\end{icompact}

\begin{figure}[t]
\centering
  \includegraphics[width=\linewidth]{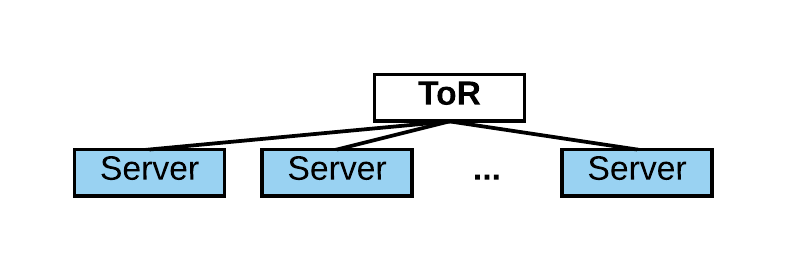}
    \vspace{-0.2in}
  \caption{Experiment Topology}\label{fig:eval:topo}
\end{figure}

\parab{Settings.} Since we do not own P4-enable switches on our testbed, in the experiments, we assume such switches are already programmed properly; if the packets are sent into the network in Scenario 2\&3, we assume that it is processed at the maximum rate (for Scenario 3, we apply a rate-limiter at each server to enforce the throughput penalty).

For the experiment hardware settings, we use 24 servers with Intel E5-2630 CPU and 32GB memory. Each server has a NetXtreme BCM5719 Gigabit Ethernet Network interface card (NIC), and are all connected to a Gigabit Ethernet switch (Figure~\ref{fig:eval:topo}).

To test raw performance and avoid unnecessary software overheads, we choose not to use existing Java-based parallel computing frameworks~\cite{hadoop, spark}. Instead, we implement a bare-bone Word-Count applications using C++. This application use the same mechanism as Map-Reduce to parallelize among multiple servers. In each server, we have a data set of a same size, one mapper, and one reducer; thus for $n$ servers, we have $n$ mappers and $n$ reducers.

\begin{figure}[t]
\centering
  \includegraphics[width=\linewidth]{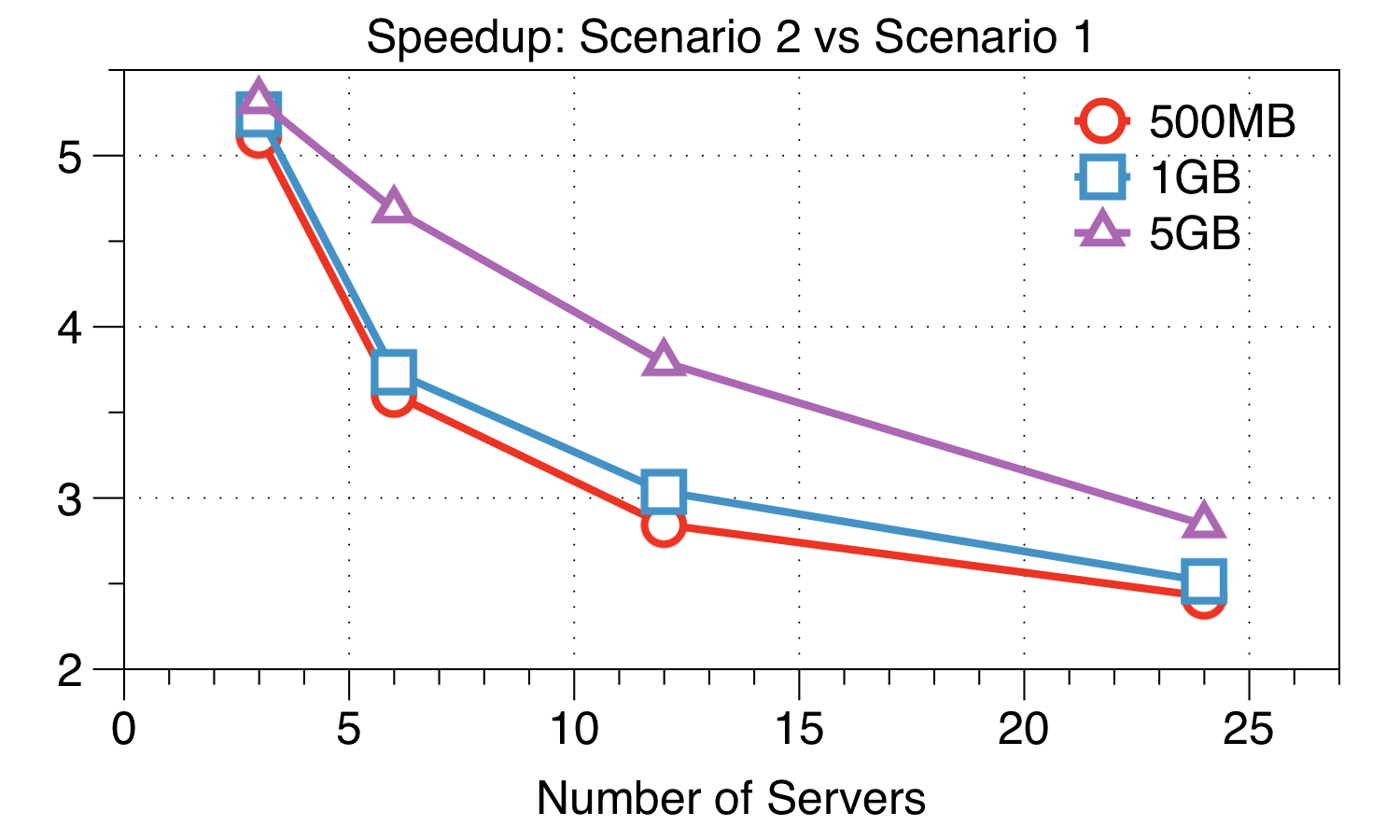}
  \caption{Experiment Results: Gain from offloading Reduce task to data plane}\label{fig:eval:21}
%\end{figure}
%
%\begin{figure}[t]
%\centering
  \includegraphics[width=\linewidth]{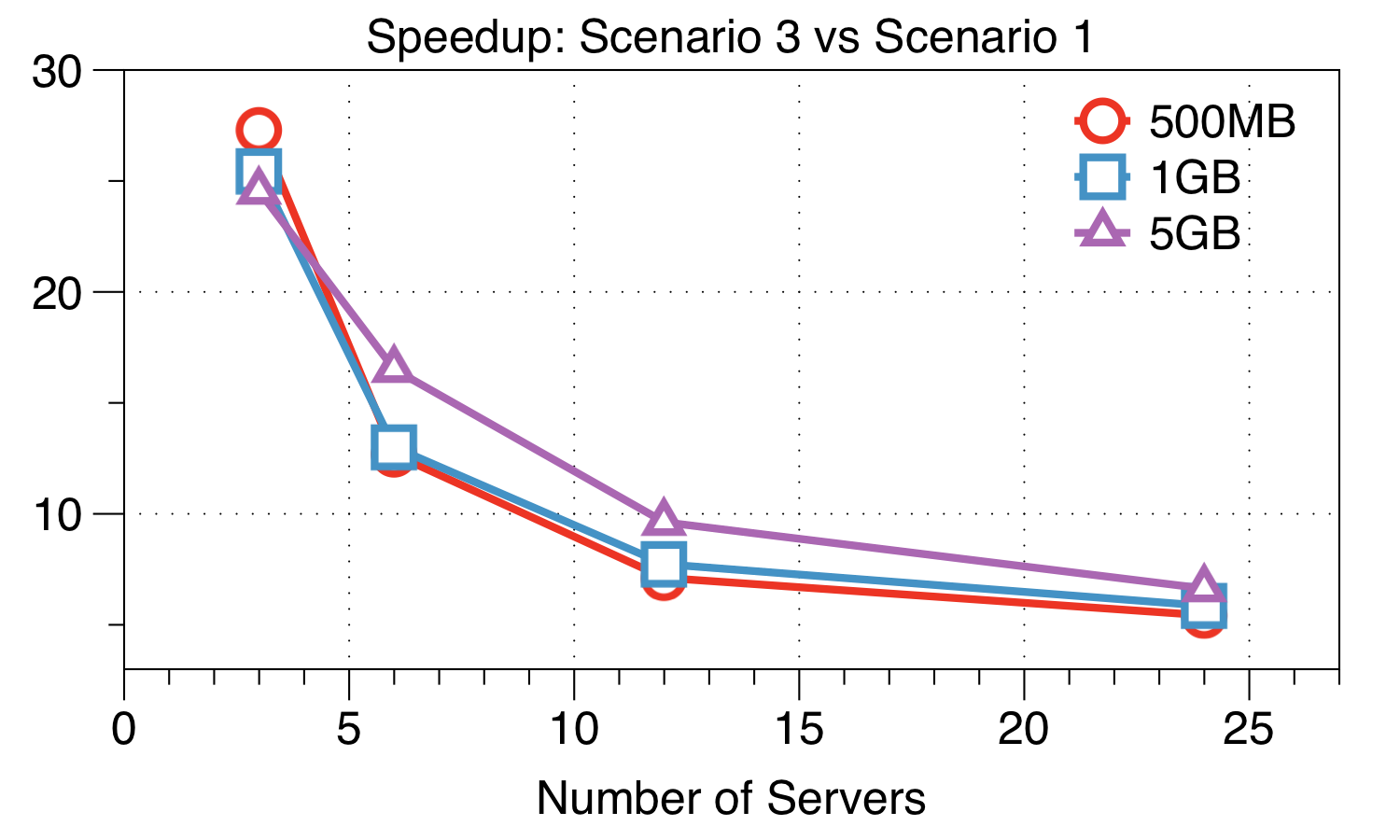}
  \caption{Experiment Results: Gain from offloading both Map and Reduce tasks to data plane}\label{fig:eval:31}
\end{figure}

\parab{Results.} We vary the total size of data set from 500MB, 1GB, to 5GB, and increase the number of servers from $3$ to $24$. The results are plotted in Figure~\ref{fig:eval:21}\&\ref{fig:eval:31}. The metric we use is job completion time (JCT) speed-up against Scenario 1, which does not use data plane computation. For each experiment, if JCT of Scenario 1 is $J_1$, and $J$ for another scenarios, than the speed-up is calculated as $J_1/J$.

\begin{figure}[t]
\centering
  \includegraphics[width=\linewidth]{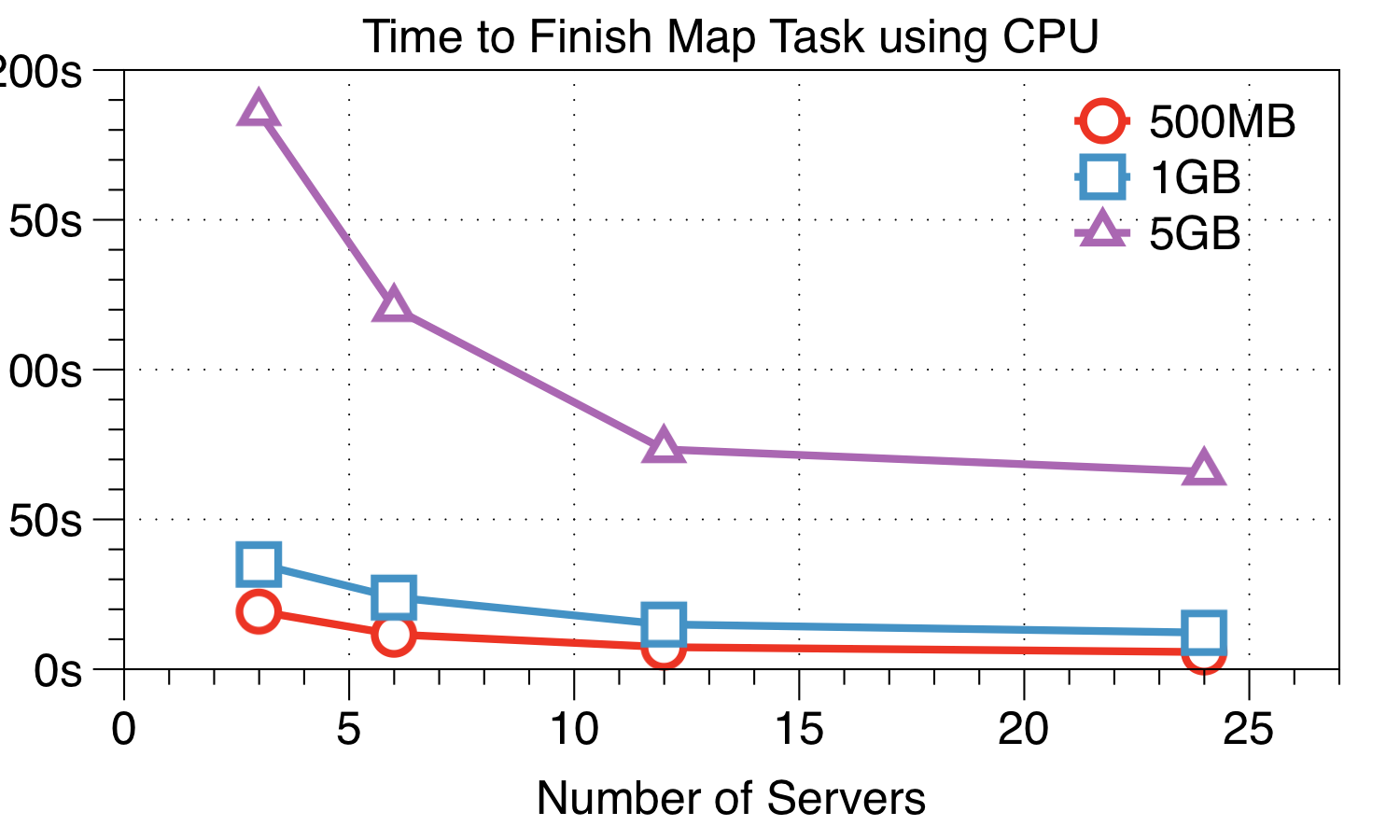}
  \caption{Experiment Results: Map Task using CPU}\label{fig:eval:map}
  \includegraphics[width=\linewidth]{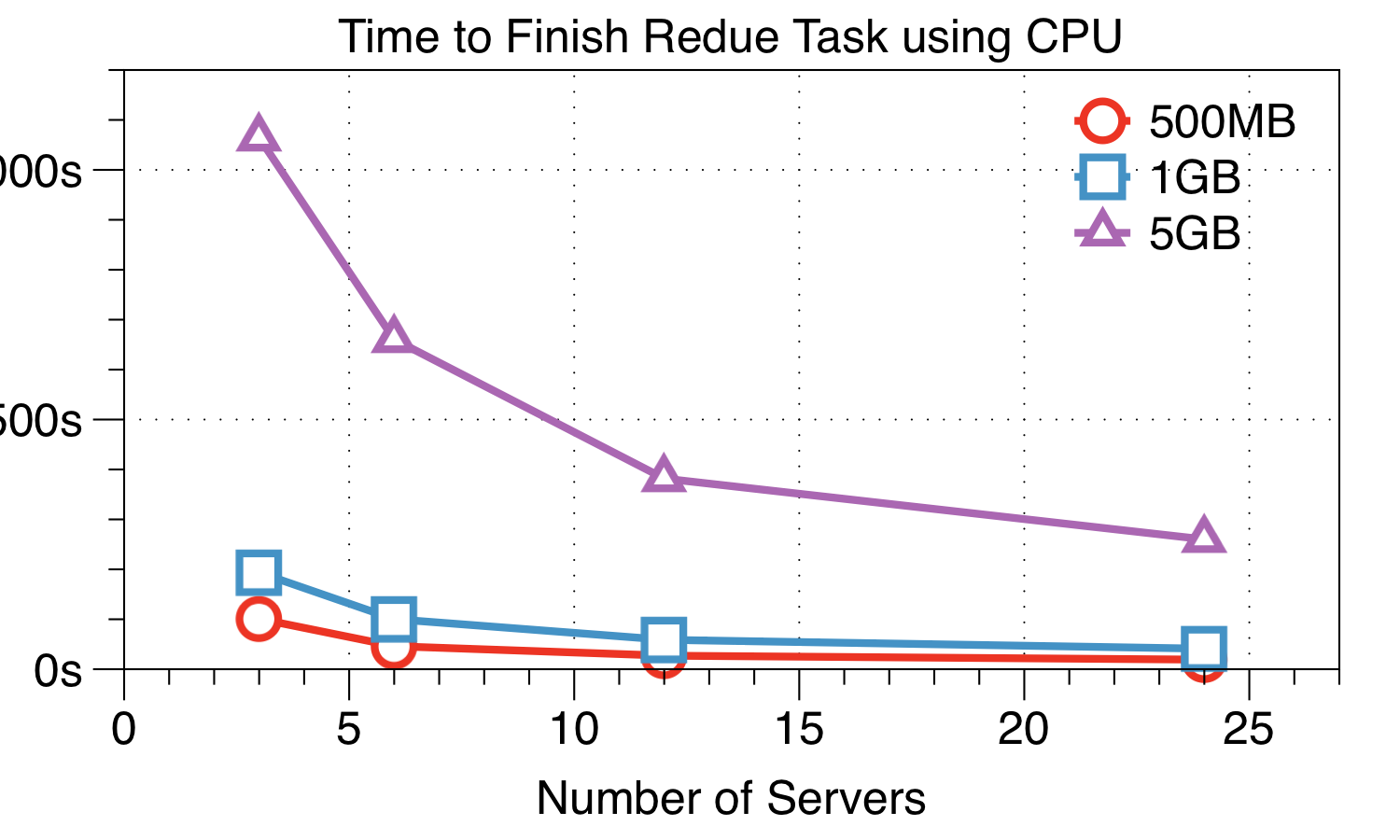}
  \caption{Experiment Results: Reduce Task using CPU}\label{fig:eval:reduce}
\end{figure}

Figure~\ref{fig:eval:21} demonstrate the gain from offloading the Reduce task to the data plane, and we observe up to $5.32\times$ speed-up. We can also see that, with larger data set, the gain is more obvious, because the larger data size requires more CPU computation for the Reduce task. Finally, with more servers added, the speed-up is decreasing. This is because the total data size is the same, and with more servers, the data size on each server is decreasing, resulting in less CPU processing on each server. As shown in Figure~\ref{fig:eval:map}\&\ref{fig:eval:reduce}, the decrease in CPU processing time for Map and Reduce task corresponds to the decrease in speed-up.

For Figure~\ref{fig:eval:31}, we can make the same observations. In addition, by comparing Figure~\ref{fig:eval:21}\&\ref{fig:eval:31}, we can see the gain from offloading the Map task to the data plane. Even with the throughput penalty (the servers sent packets at a rate of $C/e=1000Mbps/2.718=367.92Mbps$), the speed-up of Scenario 3 is at least $4.61\times$ higher than that of Scenario 2. We conclude that doing serialization of data items using CPU is much less efficient that offloading to the data plane.

In summary, we have shown with the above experiments that it is beneficial to offload some computation to the data plane.

\section{P4MR: Programming Model for Data Plane Computation}\label{sec:p4mr}
Although offloading computation to the data plane do have benefits, it is still difficult to program the multiple switches to carry out a computation task in cooperation. The users can handcraft the programs in each switch 
In this section, we propose a parallel programming framework based on P4, \sys. 

\subsection{High-Level Design}
\sys has three design goals: 1) \sys should provide primitive functions that programmable switches can support; 2) \sys should transparently parallelize user program onto multiple switches. With \sys, users can accelerate their parallel computing application by offloading certain computations to the data plane.

%\subsection{Programming models for parallel computing}
%Many programming models for parallel computing on distributed devices. We consider circuit, actor, dataflow models are suitable for a parallel programming framework for P4.

\begin{figure}[t]
\centering
  \includegraphics[width=\linewidth]{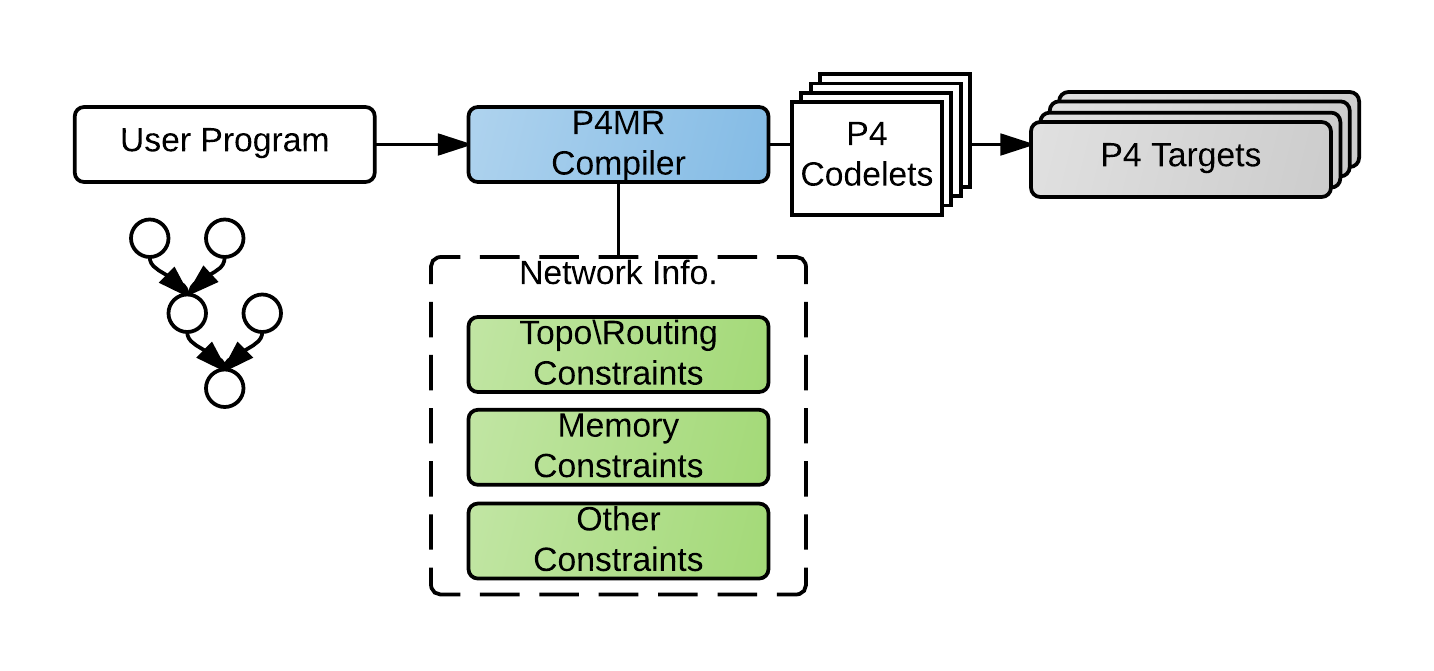}
  \caption{\sys: Overview}\label{fig:framework}
\end{figure}

\begin{figure*}[t]
\centering
  \includegraphics[width=\linewidth]{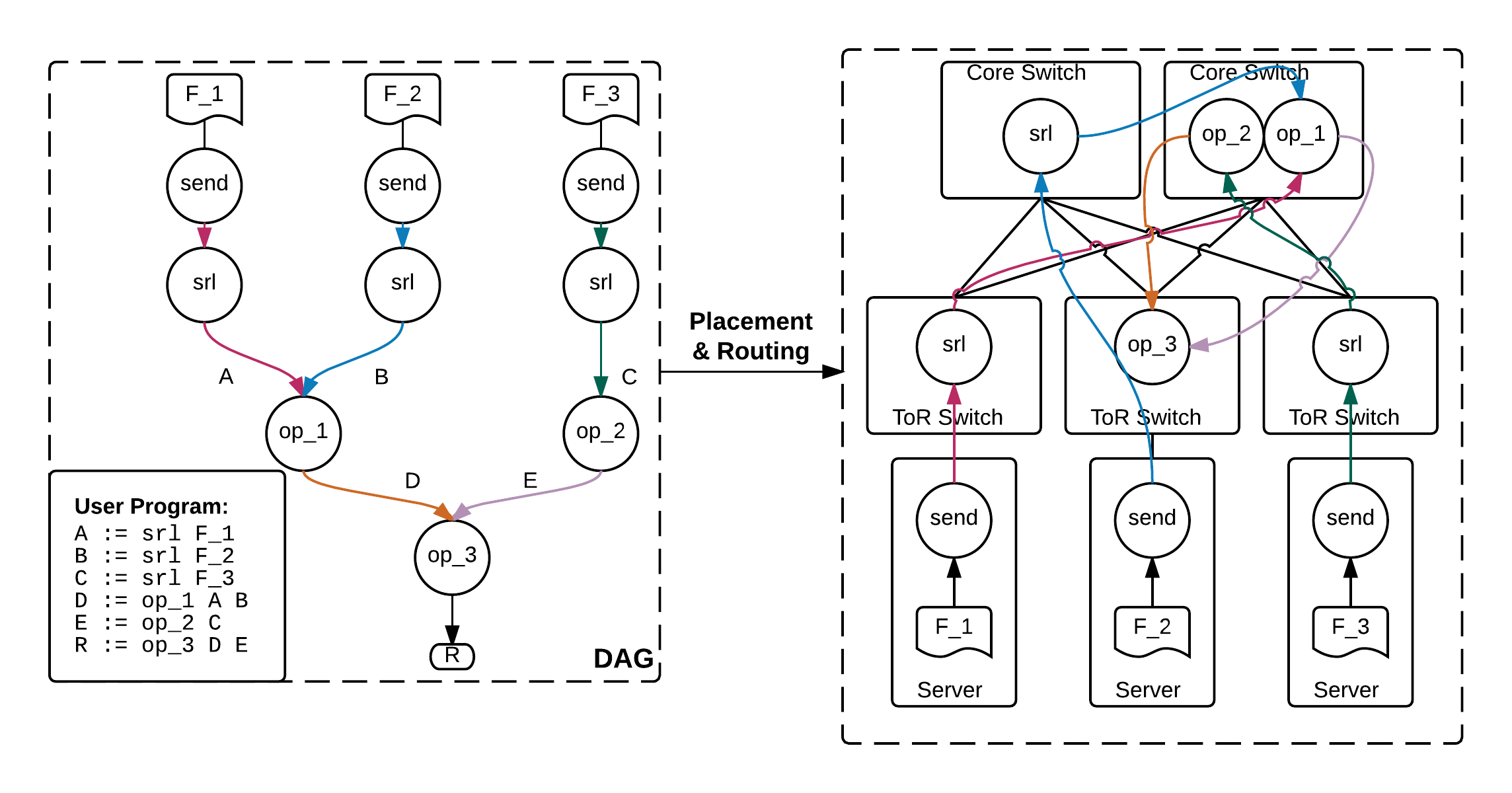}
  \caption{Compilation process}\label{fig:compile}
\end{figure*}

As depicted in Figure~\ref{fig:framework}, users compose their program using \sys-provided primitives, and submit to a cluster-wide \sys compiler. \sys compiler maintains cluster information relevant to the compilation process, such as topological, routing, and memory constraints. The compilation process follows Figure~\ref{fig:compile}: the user program is parsed into a direct acyclic graph of primitives. Given the constraints, the compiler attempts to place the primitives to the network of programmable switches. If the placement is successful, the compiler then adds appropriate routing for the packets containing data items. After placement and routing, P4 codelets for different switch in the network is generated and compiled to each switch.

\subsection{Implementation}
We have built and tested the primary model of \sys system in the open source P4
simulator \cite{p4simulator}. The simulator resides on a VMware virtual machine
based on 64-bit Ubuntu 14.04 Distribution and operates with a base memory of
2GB. The whole system consists of a \sys raw code compiler, a dependency graph
parser and finally a Mininet\cite{mininet} network simulator. In our simulation, we have mimic
a simple workflow of Map-Reduce on three files locate on different places of the
P4 Virtual Machine. Next, we are going to explain the detailed implementation of
each component with this concrete example.

To start a \sys workflow, users are required to write down their operations with
some basic MapReduce operations to explicitly define the tasks. We provide
a simple list of available operations, including \textit{assignment}, \textit{
load}, \textit{map}, \textit{summation}. Take the following code snippet as an
example: \\

\begin{minipage}{\linewidth}
%\lstset {basicstyle=\small,language=C}
\lstset {,language=C}
\begin{lstlisting}[frame=single]
A := store<uint_64>("ip_h1:path_A");
B := store<uint_64>("ip_h2:path_B");
C := store<uint_64>("ip_h3:path_C");
D := SUM(A, B);
E := SUM(C, D);
\end{lstlisting}

\end{minipage}

\begin{figure}
    \centering
    \includegraphics[width=\linewidth]{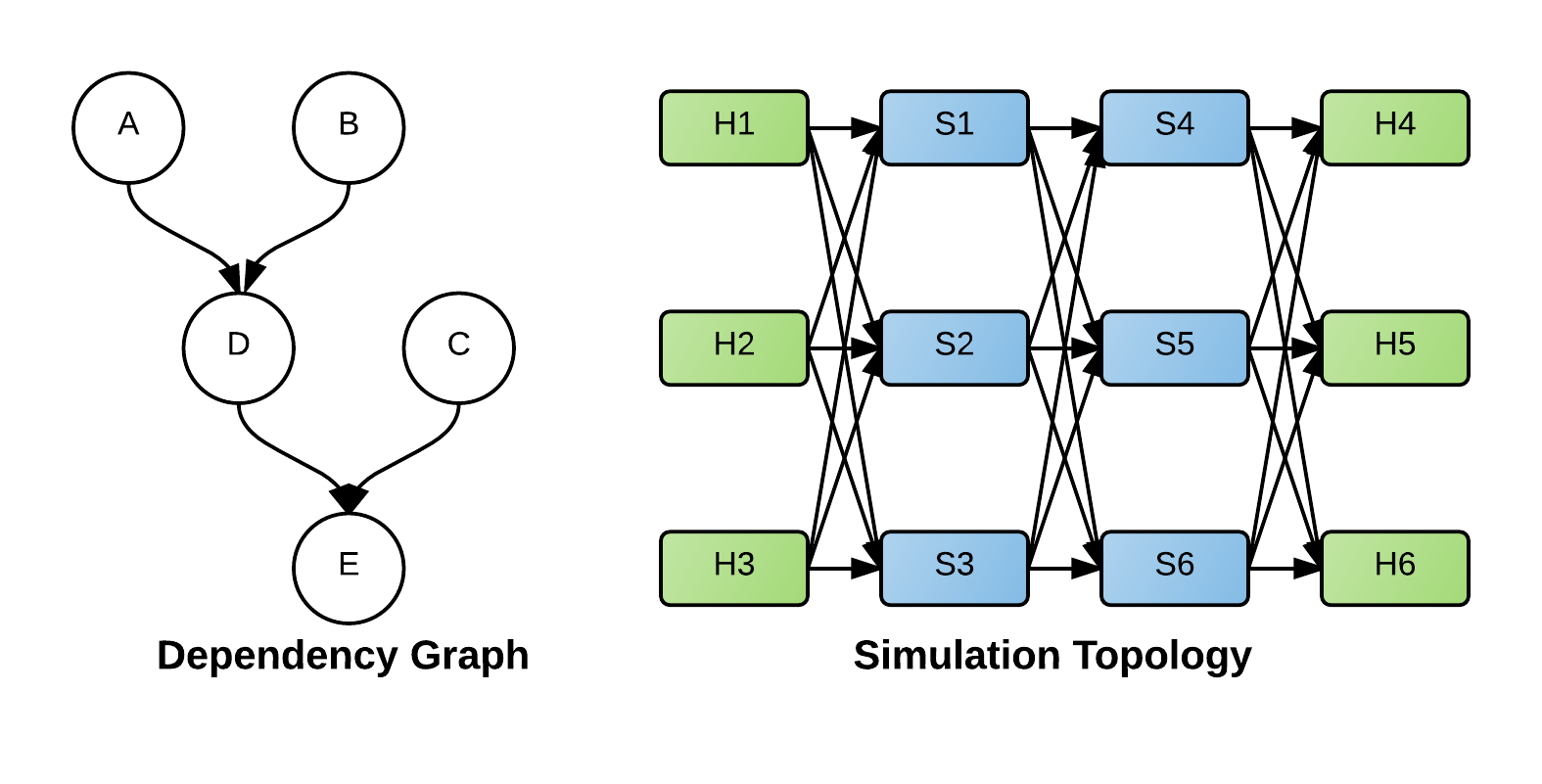}
    \caption{Implementation Example}
    \label{fig:impl}
\end{figure}

The first three lines of codes load data from different file paths to each
variable label. The last two lines specify the \textit{summation} operation
on these data source. Due to the limited operations supported by P4 Lang,
currently our simulator only support summation on 64-bit and 32-bit unsigned
integers. After the raw codes are ready, our first component, a raw code
compiler implemented with ``flex \& bison '', parses the codes and generate
an abstract syntax tree (AST) under json format. The AST contains the detailed
information of each label, such as the unique label index, function type,
parameters and so on. Later, our dependency graph parser takes the AST as
input and converts it into a directed acyclic graph (DAG). The directed acyclic
graph of the previous code listing is shown in Figure~\ref{fig:impl}.

From the dependency graph, we know that label \textit{D} depends on \textit{A}
and \textit{B} while label \textit{E} depends on \textit{D} and \textit{C}.
Based on these information, we are able to associate each label with one of
our network P4 switches. The placement of labels on switches has a significant
impact of the overall performance as it determines the routes to forward \sys
packets. The objective is to minimize the average number of hops that the whole
workflow packets will encounter. As for our preliminary design, we apply a greedy
algorithm to assign the minimum burdened switch to new labels. In our simulation,
we evaluate the previous example in the topology of Figure~\ref{fig:impl}
comprised of six hosts and six switches.

\begin{figure}
    \centering
    \includegraphics[width=0.46\textwidth]{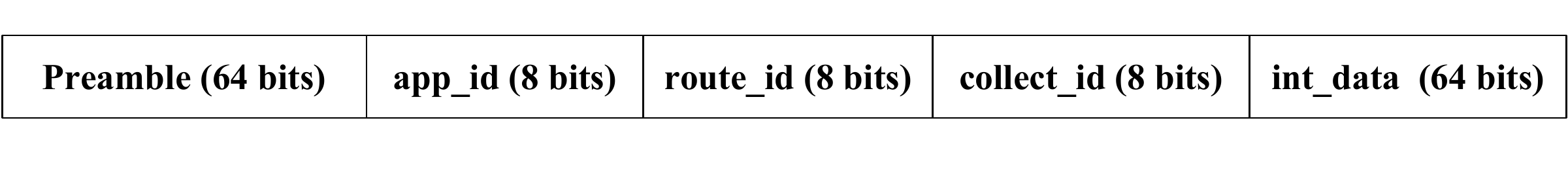}
    \caption{P4MR packet header format}
    \label{fig:pkt_format}
\end{figure}

After the label placement, the first summation labelled \textit{D} is assigned to
switch \textit{S2} and \textit{E} is associated with \textit{S6} since initially
each switch has no assignments. Once the label association is done, the Mininet
simulator in \sys will generate a routing table and reconfigure each switch from
\textit{S1} to \textit{S6} according to dependency graph. In our simulation, the
three files are initially stored on hosts \textit{h1}, \textit{h2} and \textit{h3}
respectively. Also, we randomly assign one host \textit{h6} as the results collection
end-point, onto which we forward the final results to check correctness. All the 
intermediate results are updated in the stateful variables in P4 switch. Finally,
when the \sys process is launched, packet assembler scripts segment each data
file and assemble each integer based on our pre-defined packet header format. The
packet format in our simulation is shown in Figure~\ref{fig:pkt_format}.

As described in the packet header format, a preamble field of 64 bits is used to
distinguished the P4MR traffic from others. Then, an 8-bit application id aims to 
separate different labels, collection signal or Map-Reduce functions. Next, an
8-bit routing id helps to route packets to dedicated switches. Another 8-bit
collection id is set when the current packet is a collection signal and does
not contain any valid data. Finally, a 64-bit field holds the data to update
the results. 

We have compiled and tested this prototype implementation using Mininet. The results show that this prototype is functionally correct. Since we do not have switch that support P4, we leave performance evaluation on hardware switch as future work.
\section{Conclusion}\label{sec:conclusion}
Motivated by the immense computational potential of modern programmable switches, we propose to offload some parallel computing tasks to the data plane, so that the data can be processed while in-transit. We first developed a simple mathematical model to understand the costs and overheads of data plane computation. Then we validate the the performance benefits of offloading computation to network. We then described a parallel programming framework, \sys, to help users efficiently program multiple switches. We built and tested a prototype of \sys on a simulated testbed.

\parab{Future works:} \sys still have many aspects that need improvement and further investigation.
\begin{icompact}
\item \textbf{Memory footprint.} Operational memory is one of the most precious resource on the switch, and many computational applications have large memory footprint, e.g. Word-Count. Deploying such applications on switch may use up memory and negatively impact the functionality of the switch. It is therefore important to understand the general characteristic of memory usage for each primitive functions that \sys offers to users, and to develop a predictive memory model to facilitate the compilation process.

\item \textbf{Multi-job scheduling.} Users of parallel computing frameworks expect to run multiple applications at the same time. Given multiple applications, a scheduling mechanism for \sys need to be explored.

\item \textbf{Dynamic job arrivals.} Currently \sys only support predefined user program before the network starts. When the simulated network is running, \sys cannot change the functions on the switch. Supporting dynamically arriving jobs may require incremental compilation or virtualization mechanisms\cite{hyper4}, which we intend to explore further.
  
\end{icompact}

%\clearpage
%\setlength\bibitemsep{0pt}

\bibliographystyle{IEEEtran}
\bibliography{reference}
%\clearpage
%\input{sections/appendix.tex}

%\clearpage

\end{document}